\documentclass[12pt,letterpaper]{article}

\usepackage{amsmath}
\usepackage{amsfonts}
\usepackage{amssymb}
\usepackage{graphicx}
\def\be{\begin{equation}}
\def\ee{\end{equation}}

\newcommand{\GeV}{{\rm GeV}}

\newcommand{\eV}{{\rm eV}}

\begin{document}

\begin{flushright} {\footnotesize IC/2006/004}  \end{flushright}
\vspace{5mm}
\vspace{0.5cm}
\begin{center}

\def\thefootnote{\fnsymbol{footnote}}

{\Large \bf Right-handed neutrinos as the source of density \\\vspace{.1cm} perturbations} \\[1cm]
{\large Lotfi Boubekeur and Paolo Creminelli}
\\[0.5cm]

{\small 
\textit{Abdus Salam International Center for Theoretical Physics, \\
Strada Costiera 11, 34014 Trieste, Italy}}

\end{center}

\vspace{.8cm}

\hrule \vspace{0.3cm} 
{\small  \noindent \textbf{Abstract} \\[0.3cm]
\noindent
We study the possibility that cosmological density perturbations are generated by the inhomogeneous decay of right-handed neutrinos.
This will occur if a scalar field whose fluctuations are created during inflation is coupled to the neutrino sector. Robust predictions of the model are
a detectable level of non-Gaussianity and, if standard leptogenesis is the source of the baryon asymmetry, a baryon isocurvature perturbations
at the level of the present experimental constraints.

\vspace{0.5cm}  \hrule

\def\thefootnote{\arabic{footnote}}
\setcounter{footnote}{0}

\section{Introduction}
Recently a lot of attention has been devoted to models of the early Universe where the leading source of density perturbations is not the slowly rolling inflaton but an additional light scalar. The quantum fluctuations of this additional field generated during inflation can be converted in different ways into adiabatic density perturbations. In the curvaton scenario \cite{Lyth:2001nq} the scalar starts oscillating when the Hubble scale drops below its mass and its energy density becomes quite relevant before decay. As in different parts of the Universe the initial conditions for the oscillation are slightly different this gives rise to density perturbations. Another possibility is that the scalar fluctuations modulate some parameter relevant for the cosmological evolution at some stage. For example it can modulate the reheating efficiency of the inflaton or the mass and interactions of a particle which becomes dominant before decaying \cite{Dvali:2003em,Dvali:2003ar,Kofman:2003nx}. 

These models are theoretically interesting: given the ubiquity of scalar fields in extensions of the Standard Model, it is easy to imagine that the inflaton is not the only relevant degree of freedom in the early phases of the Universe. In particular, in String Theory dimensionless couplings are functions of the scalars describing the compactification. Moreover this additional source of density perturbations can allow to circumvent the necessity of slow-roll altogether \cite{Dvali:2003vv}.  If one takes the point of view that the inflaton parameters are in some way scanned over a landscape of models, extreme values for the amplitude of density perturbations (either much smaller or much bigger than the observed $10^{-5}$) seem preferred \cite{Garriga:2005ee,Feldstein:2005bm}. This is easy to understand as changing the inflaton parameters one also changes the number of e-folds of inflation, so that the final volume exponentially depends on the parameters. If taken seriously, this can be a further motivation for looking at alternative sources of perturbations.

However the main reason behind the study of these models is that they give experimental signatures which can allow to distinguish this source of perturbations from the standard inflaton fluctuations. There are two basic observables which are relevant. First, in general these models predict a higher level of non-Gaussianity with respect to the standard slow-roll scenario, with a well defined momentum dependence. Second, given the presence of a second field, they can give rise to an isocurvature  contribution.

In an attempt to connect these ideas with realistic particle physics scenarios, in this paper we will study in detail a well defined and simple model. We will assume that the additional light scalar is coupled to right-handed (RH) neutrinos, so that their mass and decay rate are slightly different in different regions of the Universe. This will give rise to density perturbations. As we will see the production of density perturbations requires departure from thermal equilibrium. Therefore RH neutrinos appear as the most simple and minimal candidate: being SM gauge singlets, they decay out of equilibrium for sufficiently small Yukawa couplings. Fluctuations in the RH neutrino parameters will also modulate the leptogenesis process (which we assume to be the source of baryon asymmetry), thus giving rise to a baryon isocurvature contribution, correlated with the adiabatic one. We will see that the level of non-Gaussianity and the isocurvature contribution put severe constraints on the scenario. Although model dependent, the amplitude of the baryon isocurvature mode is generically predicted to be somewhat bigger than the present limits. Given this tension, the detection of baryon isocurvature should be around the corner: the model will be ruled out if the next generation experiments do not detect anything.

There are various reasons why baryon isocurvature is interesting, especially  in this class of models. First of all, given that baryon number is conserved\footnote{Given the presence of sphalerons at high temperature what we really mean is that $B-L$ is conserved.}, it is much easier to produce baryon isocurvature than for example dark matter or neutrino isocurvature. These last two are in fact erased by thermal equilibrium. 

A second point is that both the generation of the baryon asymmetry and the production of density perturbations through the modulation of some parameter are processes which require departure from thermal equilibrium.  This is well known for baryogenesis, let us see how it works for the other process. Density perturbations are produced in these models because the cosmological evolution is different in distant regions of the Universe as some relevant parameter is made space dependent by the fluctuations of the light scalar. In thermal equilibrium entropy in a comoving box is constant. This gives a unique relationship between the scale factor and the temperature of the plasma and thus it forbids the creation of density perturbations. All the explicit models are in fact based on out of equilibrium processes: reheating after inflation, inhomogeneous dominance of a massive particle (which will be the example of interest here) and one could imagine further examples like the modulation of a phase transition. Now if this out of equilibrium process is also the source of baryon asymmetry then one expects the generation of baryon isocurvature at some level. For example it is difficult to imagine that the reheating process is completely baryon symmetric\footnote{Notice that as the thermal phase of the Universe begins after the decay of the inflaton, we are not forced to look for additional sources of the baryon asymmetry. The simplest scenario is that the thermal phase just starts from an asymmetric initial condition due to a baryon asymmetric reheating. Obviously it is still interesting to try to relate the observed baryon abundance to known physics, as in the case of leptogenesis.} \cite{Murayama:1992ua}, so that some baryon isocurvature will be generated in a inhomogeneous reheating scenario. The possibility of a baryon isocurvature is well known also in the context of curvaton models \cite{Moroi:2002vx}.

Finally the baryon isocurvature experimental limits are entering now in a very interesting range from the point of view of the models we are studying. From what we said, we expect the baryon isocurvature contribution $\delta (n_B/s)/(n_B/s)$ ($n_B$ is the baryon density and $s$ the entropy density) to be comparable to the adiabatic component which can be parametrized by the usual variable $\zeta$: $\delta (n_B/s)/(n_B/s) \sim \zeta$. The most recent global data analysis \cite{Beltran:2005gr} (see also \cite{Gordon:2002gv}) puts a constraint\footnote{The constraint holds assuming that the perturbations are completely correlated or anti-correlated and have the same scale dependence. This applies to our model as both kinds of perturbations come from the same scalar $\phi$.} 
\be
\label{eq:explimit}
\left| \frac{\delta (n_B/s)}{n_B/s} /\zeta\right| < 0.7  \qquad {\rm at} \; 2 \sigma \;.
\ee
Thus we are really in a very interesting range!

In the following Section we will calculate the perturbations created by the modulation of the RH neutrino decay, focusing on two possible signatures of the model: non-Gaussianity and baryon isocurvature. We start assuming that the fluctuations in the underlying scalar field are frozen, so that their only effect is to change the neutrino parameters. We will study the scalar evolution in Section \ref{sec:phi}, before drawing our conclusions in Section \ref{sec:conclusions}.

\section{Generation of density perturbations}
In this section we want to study the production of density perturbations through the inhomogeneous dominance
and decay of right-handed neutrinos. Fluctuations in the RH neutrino parameters comes from the coupling with a scalar $\phi$, whose super-horizon perturbations are generated during inflation\footnote{Additional interactions between $\phi$ and the Standard Model are possible but irrelevant in our discussion.}. We stick to the simplest framework of Standard Model plus three hierarchical RH neutrinos. 
We concentrate for now on the decay of the lightest one and on adiabatic perturbations, postponing the 
discussion about fluctuations in the ratio $n_B/s$. We assume that this mechanism is the leading source of density perturbations; we will see
that this implies that the RH neutrino must go rather strongly out of equilibrium. If this is not the case its density remains a small fraction 
of the plasma which contains $\sim 100$ relativistic species; therefore to match the observed $ \sim 10^{-5}$ level of perturbations much 
bigger fluctuations in the decay width $\delta\Gamma/\Gamma$ and mass $\delta M/M$ are required. This would imply, as we will see explicitly below, that second 
order corrections are quite big and incompatible with the present limits on non-Gaussianities. 

The ratio $\Gamma/H$ controls whether or not the RH neutrino decays are in equilibrium. In particular it is useful to study the quantity
\be
\frac{\Gamma(T=0)}{H(T=M)} = \frac{(\lambda^\dagger \lambda)_{11} \cdot M} {8 \pi}/\Big(g_*^{1/2} \frac{M^2}{M_P}\frac{2 \,\pi^{3/2}}
{\sqrt{45}}\Big) \;,
\ee
where $M_P = G_N^{-1/2} \simeq 1.22 \times 10^{19}\,\GeV$ and $g_*$ is the number of relativistic degrees of freedom in the plasma ($g_* = 106.75$
for the Standard Model). The mass of the lightest RH neutrino is $M$ and the matrix $\lambda$ gives the Yukawa couplings of neutrinos with the Higgs.
The ratio can be rewritten introducing the Higgs VEV at low energy $v$ ($v \simeq 174\;\GeV$) in the form
\be
\frac{\Gamma(T = 0)}{H(T=M)} = \frac{\widetilde m_1}{\widetilde m^*} \;,\qquad {\rm where} \quad \widetilde m_1 \equiv 
\frac{(\lambda^\dagger \lambda)_{11} \, v^2}{M} \;.
\ee
The quantity $\widetilde m_1$ (the usual decay parameter used in leptogenesis) roughly corresponds to the contribution of the lightest RH neutrino to the left-handed neutrino mass matrix. 
On the other hand $\widetilde m^*$ is a fixed energy scale given by
\be
\widetilde m^* \equiv g_*^{1/2} \frac{v^2}{M_P} \frac{16\,\pi^{5/2}}{3 \sqrt{5}} \simeq 1.1 \times 10^{-3} \,\eV \;.
\ee
For $\widetilde m_1 \ll \widetilde m^*$ RH neutrinos are out of equilibrium and decay at a temperature much smaller than their mass. If the decay happens very late
neutrinos dominate the Universe before decaying. This will happen after the Universe has expanded by a factor $g_*$ since $T \sim M$.
As in radiation dominance $H \propto a^{-2}$ we need $\widetilde m_1 \ll \widetilde m^*/g_*^2$. 

We are interested in the density perturbations induced by the inhomogeneous dominance and decay of RH neutrinos; we assume that RH neutrinos start from a thermal distribution for $T \gg M$. In the absence of additional sources
of perturbations, at temperature $T \gg M$ the Universe is unperturbed and the surfaces of constant temperature are flat. Different
regions of the Universe will be characterized by different values of the scalar $\phi$, so that the mass and decay width of the 
RH neutrino will be different. As a consequence the neutrino will become non-relativistic and decay at slightly different temperatures in different 
parts of the Universe. If we take two reference temperatures, one above the RH neutrino mass (let us call it $T_{\rm high}$) and one much below, after the 
neutrino decay ($T_{\rm low}$), the ratio of the scale factors between these two temperatures $a(T_{\rm low})/a(T_{\rm high})$ 
will not be exactly the same in different parts of the Universe. Thus if the surface of constant $T_{\rm high}$ is flat, 
the one of $T_{\rm low}$ will be curved. Modes of cosmological interest are huge compared to the horizon at the temperature we are looking at. 
In this limit adiabatic scalar perturbations can be expressed, in coordinates in which the surfaces of constant time are of constant temperature, as
\be
ds^2 = - dt^2 + e^{2 \zeta(\vec x)} a(t)^2 d \vec x^2 \;.
\ee
We have neglected terms subleading in the expansion $k/(aH)$, where $k$ is the comoving wavevector of the perturbation. In this limit the variable 
$\zeta$ is constant in time at any order of perturbation theory. The reason for this is quite clear: $\zeta$ just describes a locally unobservable 
rescaling of the spatial coordinates \cite{Salopek:1990jq,Maldacena:2002vr}. From the discussion above we conclude that perturbations induced 
by the RH neutrino can be obtained by the equality
\be
\label{eq:2zeta}
e^{\zeta(\vec x)} = \frac{a(T_{\rm low})}{a(T_{\rm high})} (M, \Gamma) \;.
\ee
Spatial fluctuations of $M$ and $\Gamma$ induce fluctuations in $\zeta$. In this way the problem of calculating density perturbations is reduced
to evaluate the right hand side of the equation above in an {\em unperturbed} Universe as a function of the parameters describing the RH neutrino.

Let us start from the simple case in which the RH neutrino dominates before decaying. As we discussed this will happen for 
$\widetilde m_1 \ll \widetilde m^*/g_*^2$. In this case we can divide the evolution of the Universe around this epoch into three periods
with different equation of state. (1) At temperatures $T \gg M/g_*$ the Universe is radiation dominated. (2) The Universe is dominated by 
the RH neutrino (matter dominance) from $T \simeq  M/g_*$ until its decay at $H \sim \Gamma$. (3) After the neutrino decays we are back in radiation
dominance. It is easy to calculate the expansion in the three stages
\begin{eqnarray}
\frac{a(T \simeq M/g_*)}{a(T_{\rm high})} & \propto & M^{-1} \\
\frac{a(H \simeq \Gamma)}{a(T \simeq M/g_*)} & \propto & \left(\frac{H(T \simeq M/g_*)}{\Gamma}\right)^{2/3} \propto M^{4/3} \,\Gamma^{-2/3} \\
\frac{a(T_{\rm low})}{a(H \simeq \Gamma)} & \propto & \left(\frac{\Gamma}{H(T_{\rm low})}\right)^{1/2} \propto \Gamma^{1/2} \;.
\end{eqnarray}
We can put these results together to get \cite{Dvali:2003ar}
\be
\label{eq:analytic}
\frac{a(T_{\rm low})}{a(T_{\rm high})} \propto M^{1/3} \, \Gamma^{-1/6} \propto \widetilde m_1^{-1/6}  \qquad {\rm for} \quad 
\widetilde m_1 \ll \widetilde m^*/g_*^2\;.
\ee
Note that we are not interested in the proportionality factor as it corresponds to a space independent rescaling of $a$. In the limit
we studied the result just depends on $\widetilde m_1$ and not on $M$ and $\Gamma$ independently. It is easy to see that this holds
in general. If we change $M$ keeping $\widetilde m_1$ fixed we are moving the window of temperature where the equation of state deviates
significantly from radiation dominance, but this does not change the expansion between the reference temperatures $T_{\rm high}$ and $T_{\rm low}$.
In fact, once we fix $\widetilde m_1$, the scale factor becomes a function of the ratio $T/M$ only and not of $T$ and $M$ independently; as both at high and at low temperature we are in 
radiation dominance the dependence on $M$ simplifies in the ratio $a(T_{\rm low}/M)/a(T_{\rm high}/M)$. Thus relation (\ref{eq:2zeta})
simplifies to 
\be
\label{eq:2zeta2}
e^{\zeta(\vec x)} = \frac{a(T_{\rm low})}{a(T_{\rm high})} (\widetilde m_1) \;.
\ee

Let us now relax the inequality $\widetilde m_1 \ll \widetilde m^*/g_*^2$. For larger values of $\widetilde m_1$ the RH neutrino does not 
dominate the plasma before decay; increasing $\widetilde m_1$ the deviation from simple radiation dominance becomes smaller and smaller until
eventually the ratio of scale factors $a(T_{\rm low})/a(T_{\rm high})$ approaches the constant value 
$T_{\rm high}/T_{\rm low}$.
The solution for a generic $\widetilde m_1$ can be easily obtained numerically by solving the coupled equations describing the decay of the 
RH neutrino fluid into radiation
\begin{eqnarray}
\dot\rho_\gamma &+& 4 H \rho_\gamma = \Gamma \rho_N \\
\dot\rho_N  &+& 3 \big(1+w_N(T/M)\big) H \rho_N = -\Gamma \rho_N \\
H^2 &=& \frac{8 \pi}{3 M_P^2} (\rho_N + \rho_\gamma) \;.
\end{eqnarray} 
We are assuming that the RH neutrino has a thermal population at very high temperature so that its energy density $\rho_N$ starts
as $1/g_*$ of the one in the plasma $\rho_\gamma$. The pressure to energy density ratio $w_N = p_N/\rho_N$ of the neutrinos drops from $1/3$ to $0$
as they become non-relativistic at $T \sim M$. We will see that the decay must occur sufficiently out of equilibrium to be consistent with
the limits on non-Gaussianities, so it is self-consistent to neglect in the equations the inverse decay of the RH neutrino. We also neglect 
$\Delta L =1$ reactions which give contributions suppressed by a loop factor and thermal corrections, which are small as the decay occurs at 
$T \ll M$ \cite{Giudice:2003jh}. In Fig.~\ref{fig:expansion} we plot the ratio $a(T_{\rm low})/a(T_{\rm high})$ as a function of the 
parameter $\widetilde m_1$. For $\widetilde m_1 \ll \widetilde m^*/g_*^2$ we recover the analytic result of eq.~(\ref{eq:analytic}); while
for much bigger $\widetilde m_1$ we approach the radiation dominance result.

\begin{figure}[th!!]             
\begin{center}
\includegraphics[width=14cm]{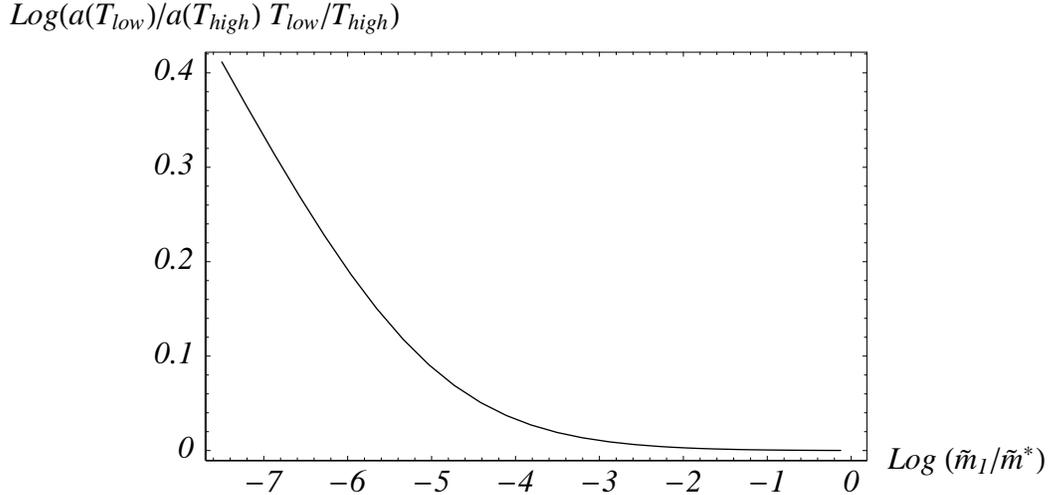}
\caption{\label{fig:expansion} \small Expansion between two fixed temperatures as a function of  $\widetilde m_1$.}
\end{center}
\end{figure} 

\subsection{Non-Gaussianity}
Now that we have calculated the expansion ratio as a function of $\widetilde m_1$, we are ready to evaluate the amount of 
non-Gaussianity. Relation (\ref{eq:2zeta2}) between $\widetilde m_1$ and $\zeta$ is non-linear and this is the source of non-Gaussianity.
Let us assume for the moment that the ratio $\delta\widetilde m_1/\widetilde m_1$ is distributed as a Gaussian. Expression (\ref{eq:2zeta2}) 
gives $\zeta$ in terms of $\widetilde m_1$ {\em at the same point in space}. In this case, the form of non-Gaussianity induced by second order terms 
is usually called {\em local} 
\footnote{The shape dependence of the 3-point function in different models has been studied in \cite{Babich:2004gb,Creminelli:2004yq}.} and parametrized 
by the parameter $f_{\rm NL}$ defined by  
\be
\label{eq:fNL}
\zeta(\vec x) = \zeta_g(\vec x) -\frac35 f_{\rm NL} (\zeta_g(\vec x)^2- \langle \zeta_g^2\rangle) \;,
\ee
where $\zeta_g$ is a Gaussian random variable.

From the function $\zeta = f(\log \widetilde m_1/\widetilde m^*)$ shown in Fig.~\ref{fig:expansion}, we have at second order
\be
\label{eq:exp}
\zeta (\vec x) = f' \frac{\delta\widetilde m_1}{\widetilde m_1}(\vec x) + 
\frac12 (f''-f') \left(\frac{\delta\widetilde m_1}{\widetilde m_1}(\vec x)\right)^2 \;,
\ee
where we have subtracted an irrelevant constant value in $\zeta$. From the linear term we get the relationship between the spectrum of 
$\delta\widetilde m_1/\widetilde m_1$ and the one of $\zeta$, while the second order term gives
\be
\label{eq:fNLf}
f_{\rm NL} = -\frac56 \frac{f''-f'}{f'^2} \;.
\ee
As we expected this expression becomes very large for large $\widetilde m_1$. The RH neutrino becomes less and less relevant at decay, so that keeping
fixed the observed value $10^{-5}$ for density perturbations, we have to correspondingly increase the fluctuations $\delta\widetilde m_1/\widetilde m_1$.
This implies that second order corrections become more and more relevant. This is an example of a quite generic correlation between the``inefficiency" of the mechanism producing density perturbations and the level of non-Gaussianity \cite{Zaldarriaga:2003my}.  The non-linearity parameter $f_{\rm NL}$ will be of the order of the inverse of the fraction of energy density in the RH neutrino at decay:
\be
|f_{\rm NL}| \sim g_* \sqrt\frac{\widetilde m_1}{\widetilde m^*} \;.
\ee
In Fig.~\ref{fig:fnl} we plot $f_{\rm NL}$ as a function of $\widetilde m_1$ from eq.~(\ref{eq:fNLf}).
\begin{figure}[th!!]             
\begin{center}
\vspace{0.3cm}
\includegraphics[width=11.5cm]{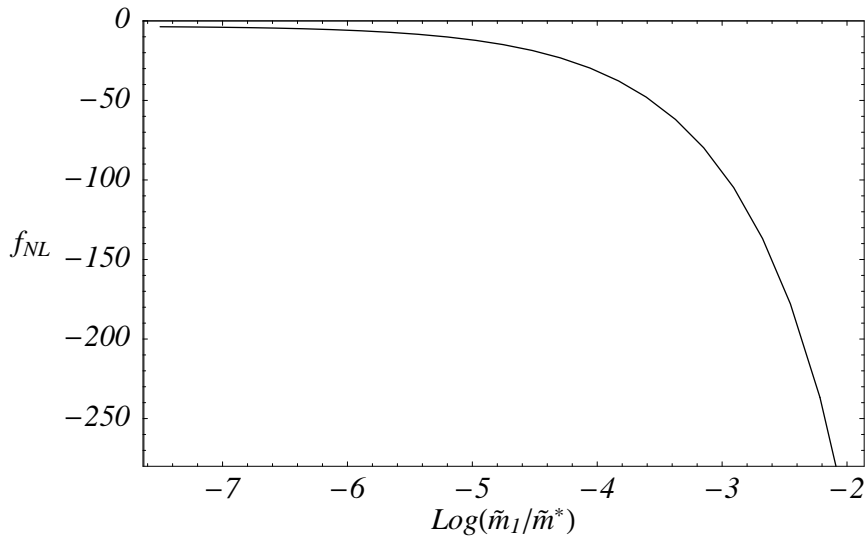}
\caption{\label{fig:fnl} \small The non-Gaussianity parameter $f_{\rm NL}$ as a function of $\widetilde m_1$.}
\end{center}
\end{figure} 

In the discussion we assumed that the ratio $\delta\widetilde m_1/\widetilde m_1$ is distributed as a Gaussian. This will not be the case in general;
the dependence $\widetilde m_1(\phi)$ will induce additional non-linearities similarly to what happened in the relation $\zeta(\widetilde m_1)$. It is
easy to realize that also these non-linearities will be enhanced by the ``inefficiency'' of the mechanism when the RH neutrino does not dominate.
The coefficient $f'$ in eq.~(\ref{eq:exp}) becomes small in this limit and second order terms in the expansion of $\widetilde m_1(\phi)$ give a 
contribution to $f_{\rm NL}$ enhanced by $f'^{-1}$. We conclude that although the plot of Fig.~\ref{fig:fnl} should not be taken as a sharp 
prediction we do not expect, barring unlikely cancellations, a smaller value for $f_{\rm NL}$ (\footnote{Note that even the negative sign for 
$f_{\rm NL}$ in Fig.~\ref{fig:fnl} can be flipped by second order effects from $\widetilde m_1(\phi)$.}). 

In addition to the non-linearities we discussed the statistics of $\phi$ itself may be non-Gaussian as a consequence of its self-interactions. The leading effect will be given by the non-linear superhorizon evolution, which gives a non-Gaussianity of the form (\ref{eq:fNL}) \cite{Zaldarriaga:2003my}. Also this contribution is quite model dependent and again enhanced by the ``inefficiency'' $f'^{-1}$.

The most stringent experimental limits on the parameter $f_{\rm NL}$ come from the analysis of WMAP data \cite{Komatsu:2003fd,Creminelli:2005hu}.
The allowed $2 \sigma$ range $-27 < f_{\rm NL} < 121$ can be converted into an approximate (given the unknown contributions discussed above) 
bound on $\widetilde m_1$
\be
\label{eq:limit}
\widetilde m_1 \lesssim 10^{-3} \cdot \widetilde m^* \simeq 10^{-6} \eV \;.
\ee
This means that the RH neutrino Yukawa interactions are strongly out of equilibrium at decay. This implies that the same interactions cannot 
bring the neutrinos in thermal equilibrium at $T \gg M$; we will come back to this point in the following Section. 

Notice that in general inequality (\ref{eq:limit}) has no direct implication for the low energy neutrino parameters as
the neutrino parameters relevant for the production of density perturbations are different from the ones measured at low energy. Indeed the scalar
starts displaced from the origin where it will be eventually stabilized. If one assumes that the variation of parameters is not very large, the inequality above implies that one of eigenvalues of the 
left-handed neutrino mass matrix is smaller than $10^{-6}\;\eV$. In fact it is easy to prove that $\widetilde m_1$ cannot be smaller than all the 
three neutrino mass eigenvalues \cite{Fujii:2002jw}. From the constraints we have from neutrino oscillation experiments, this state would be the lightest 
of the three, with a resulting hierarchical spectrum.

\subsection{Baryon isocurvature}
We now come to the discussion of the baryon isocurvature contribution. We will  see that this puts additional severe bounds on the model. The ratio between baryon and entropy density produced by RH neutrino decay after reprocessing by sphaleron transitions can be written as \cite{Giudice:2003jh}
\be
\frac{n_B}{s} = -\frac{28}{79} \; \epsilon_{N_1} \; \eta(\widetilde m_1) \; \frac{n_{N_1}}{s}(T \gg M) \;.
\ee
The factor $\epsilon_{N_1}$ is the CP-asymmetry parameter in $N_1$ decay which is given by
\be
\label{eq:CP}
\epsilon_{N_1} = \frac1{8 \pi} \sum_{j \neq 1} \frac{{\rm Im}(\lambda^\dagger\lambda)^2_{j1}}{(\lambda^\dagger\lambda)_{11}} 
f\left(\frac{m_{N_j}^2}{m_{N_1}^2}\right) 
\ee
\be
f(x) = \sqrt{x} \left[\frac{x-2}{x-1}-(1+x)\log\left(\frac{1+x}{x}\right)\right] \stackrel{x \gg 1}{\longrightarrow} -\frac3{2\sqrt{x}} \;.
\ee
The ratio $n_{N_1}/s$ is calculated in thermal equilibrium at high temperature and it is given by $n_{N_1}/s = 135 \,\zeta(3)/(4 \pi^4 g_*)$. Finally $\eta$ is an
efficiency parameter\footnote{The efficiency parameter depends solely on $\widetilde m_1$, unless the mass of the RH neutrino is very large: 
$M \gtrsim 10^{13}\GeV$ \cite{Giudice:2003jh}.} which takes into account the wash-out of the produced baryon asymmetry (which is relevant if the RH neutrino decays close to 
equilibrium $\widetilde m_1 \gtrsim \widetilde m^*$) and the entropy released in the decay (which is relevant when the RH neutrino
becomes dominant in the Universe $\widetilde m_1 \lesssim \widetilde m^*/g_*^2$).

Let us start assuming that the CP violating parameter $\epsilon_{N_1}$ does not depend on the scalar $\phi$ so that it is constant in space. Given the constraints we discussed from non-Gaussianities, the decay of the RH neutrino must be strongly out of equilibrium so that the wash-out of the produced baryon asymmetry can be neglected. Every neutrino becomes (on average) a constant fraction $-28/79 \,\cdot \epsilon_{N_1}$ of a baryon, independently of $\widetilde m_1$. This implies that the surfaces of constant baryon density remain flat while, as we discussed, the surfaces of constant temperature become curved after the RH neutrino decay. This mismatch gives a baryon isocurvature 
component
\be
\label{eq:fuck3}
\frac{\delta (n_B/s)}{n_B/s}  = - \frac{\delta s}{s} = -3 \frac{\delta T}{T} = -3 \zeta \;.
\ee
This isocurvature component is too large and it is ruled out by the present limits in (\ref{eq:explimit}). Additional contributions
will come from fluctuations in the CP violating parameter $\epsilon_{N_1}$: if the scalar $\phi$ enters in the mass term and/or in the Yukawa
interactions of the RH neutrinos this will induce fluctuations in $\epsilon_{N_1}$. Thus we have
\be
\frac{\delta (n_B/s)}{n_B/s}/\zeta  = -3 + \frac{\delta \epsilon_{N_1}/\epsilon_{N_1}}{\zeta} \;.
\ee
Roughly we expect also this second contribution to give an order one isocurvature component. From the explicit expression for $\epsilon_{N_1}$ eq.~(\ref{eq:CP}), we see that the precise number is very model dependent as it depends on how the scalar $\phi$ enters in the different
entries of the Yukawa matrix and in the right-handed neutrino masses.

We conclude that a moderate cancellation between the two terms above is required to be consistent with the present limits. From a more optimistic point of view we can say that detection of an isocurvature component should be around the corner, as it is naturally expected to be of order one in these models.

What happens if we consider not only the lightest RH neutrino but also the other two (assuming that they all start with thermal abundance at high temperature)? We can study two interesting cases. In the case all the three neutrinos are not in thermal equilibrium when they become non-relativistic ($\widetilde m_i \ll \widetilde m^*$ for all $i$), then the situation is pretty similar to the case studied below with just a single RH neutrino. All the three neutrinos will contribute to the generation of $\zeta$  (we can take the reference temperatures $T_{\rm high}$ and
$T_{\rm low}$ to encompass all the neutrinos) and to the baryon asymmetry. Note that as all the neutrinos are out of equilibrium the baryon asymmetry generated by a RH neutrino is not washed out by the lighter ones. Again if we take all the CP violating parameters $\epsilon_i$ as constant we get an isocurvature component 3 times bigger than the adiabatic one as in eq.~(\ref{eq:fuck3}). Thus again a moderate cancellation between this contribution and the one generated by $\delta\epsilon_i/\epsilon_i$ is required.

A quite different situation is the case in which the lightest RH neutrino $N_1$ is close to equilibrium when it becomes non-relativistic. As we discussed its contribution to $\zeta$ will be small in this case and can be neglected. As $N_1$ is in equilibrium after the heaviest neutrinos
decay, it will wash-out the lepton asymmetry already produced. Although this wash-out effect depends on the flavor structure and can be not entirely efficient \cite{Barbieri:1999ma,Vives:2005ra}, we can assume for simplicity that all the baryon asymmetry is generated by the lightest RH neutrino. Thus now the two mechanisms are separated: fluctuations in the parameters of the heavy $N_{2,3}$ generate $\zeta$ while fluctuations in the decay of $N_1$ give baryon isocurvature. In the limit in which the scalar $\phi$ is decoupled from $N_1$ ({\em i.e.} it does not change neither its mass nor its couplings) we have no isocurvature. As $N_1$ is close to equilibrium the wash-out of the produced lepton asymmetry cannot be neglected. A good parameterization of the efficiency parameter $\eta(\widetilde m_1)$ in the regime $\widetilde m_1 \gg \widetilde m^*$ is given by $\eta \propto \widetilde m_1^{-1.16}$ \cite{Giudice:2003jh}.
Fluctuations in this efficiency parameter and in $\epsilon_{N_1}$ will give an isocurvature component. Thus, although in this case we do not have to rely on a cancellation to be compatible with data, we still expect the isocurvature contribution to be of order one, unless for some particular reason the coupling of $\phi$ with $N_1$ is suppressed.

More complicated intermediate scenarios are possible but, without further specifying the coupling of $\phi$, the generic conclusion we can draw is that the isocurvature contribution should be around the (experimental) corner.    

Up to now we assumed that, at high temperatures, RH neutrinos are in thermal equilibrium. On the other hand we need that, at least for one of the neutrinos,
the decay parameter $\widetilde m$ is sufficiently small to allow the RH neutrino to become a significant fraction of the plasma before decay. This also
implies that Yukawa interactions are not fast enough to bring the neutrino in thermal equilibrium starting from a generic initial condition. Additional 
interactions are required to produce neutrinos at high temperature; for example gauge interactions mediated by a heavy GUT gauge boson 
\cite{Plumacher:1996kc}. 
If there are no additional interactions we expect RH neutrinos to get produced at some level by the reheating process, either perturbatively or by 
their coupling with the coherent inflaton oscillations \cite{Giudice:1999fb}. In this case the initial RH neutrino density is model 
dependent and we do not even know their
pressure to energy ratio \cite{Podolsky:2005bw}. The analysis above obviously does not apply to this case, although we still expect a significant baryon isocurvature component
as the generation of density perturbations and baryogenesis are unified in the same process. We can think about this non-thermal production of RH neutrinos
as part of the reheating process, before getting to a completely thermal state. From this point of view the modulated decay of RH neutrinos becomes 
a modulated reheating scenario \cite{Dvali:2003em,Kofman:2003nx}. This is a nice example of a more general point: if the baryon asymmetry is produced at reheating 
(which is the simplest mechanism for baryogenesis), we expect that models with modulated reheating will give a baryon component.

\section{\label{sec:phi}The evolution of the scalar $\phi$}
So far we concentrated on the effects of the modulation of the RH neutrino parameters induced by the fluctuations of a scalar $\phi$.
Another aspect one should consider is the effect of the scalar couplings with the neutrino sector on the dynamics of the scalar itself.  We have tacitly assumed that the scalar does not evolve in time around the epoch when RH neutrinos become non-relativistic, so that its only effect is to make the parameters space dependent. We can now ask if this is a good assumption. Let us start assuming that $\phi$ enters only 
in the RH neutrino mass matrix with a coupling (neglecting all the flavor structure)  
\be
\frac12 M(\phi/M_P) NN \;.
\ee
Without loss of generality we choose to normalize the scalar with $M_P$ in the function $M$. The scalar is coupled to the plasma of RH neutrinos;  assuming that they start with a thermal distribution we have $\langle N N\rangle \sim T^3$, so that the evolution of $\phi$ will satisfy
\be
\ddot\phi + 3 H \dot\phi + \frac{M'}{M_P} T^3 =0 \;.
\ee
We are assuming that the bare potential for $\phi$, $V(\phi)$, is sufficiently flat to be irrelevant at the epoch we are looking at\footnote{This requirement limits the amount of non-Gaussianity coming from the self-interactions of $\phi$ \cite{Zaldarriaga:2003my}.}.
The variation of $\phi$ during an Hubble time will thus be of order
\be
\Delta\phi \sim \frac{M'}{M} \frac {M T^3}{H^2 M_P^2} M_P \;.
\ee
As we need the RH neutrino to become dominant to be a sufficiently efficient source of density perturbations, during dominance 
we have $M T^3/(H^2 M_P^2) \sim 1 $ which gives
\be
\Delta\phi \gtrsim \frac{M'}{M} M_P \;.
\ee 
On the other hand the induced level of density perturbations will be at most of order
\be
\zeta \sim \frac{M'}{M} \frac{\delta\phi}{M_P} \sim \frac{M'}{M} \frac{H_{\rm inflation}}{M_P} \;.
\ee
From the fact that we did not observe gravitational waves created during inflation, we have an upper bound on $H_{\rm inflation}$ which, given the observed value of $\zeta$, gives a lower bound on $M'/M$ of order $M'/M \gtrsim 10$ (\footnote{Notice that we are not using the constraint \cite{Dvali:2003ar,Vernizzi:2005fx} that the level of density perturbations created by our mechanism is dominant with respect to the standard contribution from the inflaton $\zeta \sim H/(M_P \sqrt{\epsilon})$. This constraint (which would give a lower bound on $M'/M$ a bit stronger than the one above) could be circumvented in models of inflation without slow-roll \cite{Dvali:2003vv}. On the other hand the gravitational wave constraint discussed in the text is completely general.}). This implies that the scalar moves by more than $M_P$ during the epoch we are interested in! Even taking the natural range for $\phi$ of order $M_P$, this motion will induce a very big variation in the neutrino
parameters. This motion makes all the analysis more complicated and dependent on the stabilization potential of the scalar.

The situation is much more under control in the case the scalar enters only in the Yukawa couplings of RH neutrinos
\be
\lambda (\phi/M_P) L H N \;.
\ee
In this case, as $\langle L H N \rangle \sim \lambda T^4$, we have an equation of motion
\be
\ddot\phi + 3 H \dot\phi + \frac{\lambda' \lambda}{M_P} T^4 =0 \;.
\ee
The variation of $\phi$ in one Hubble time is thus
\be
\Delta\phi \sim \frac{\lambda'}{\lambda} \frac {\lambda^2 T^4}{H^2 M_P^2} M_P \;.
\ee
while the expression for density perturbations is 
\be
\zeta \sim \frac{\lambda'}{\lambda} \frac{H_{\rm inflation}}{M_P} \;.
\ee
We see that, contrary to the previous case of mass variation, for small enough $\lambda$ the displacement of the scalar is negligible without suppression of the produced density perturbations.  We conclude that this scenario with fluctuating Yukawa couplings is favored with respect to mass fluctuations.

We have discussed the evolution of $\phi$ at the time density perturbations are generated, but what happens to it after that? Eventually
the mass term in the scalar potential $V(\phi)$ will get larger than the Hubble rate $H$. At this point the scalar starts oscillating around the minimum and
the energy stored in these oscillations red-shifts as $a^{-3}$, {\em i.e.} slower than radiation. If the displacement of the scalar from the minimum is 
of order $M_P$, the oscillations start dominating soon after $m_\phi \gtrsim H$. In this case the scalar would act as a curvaton \cite{Lyth:2001nq}: 
everything created before gets redshifted away (including the generated baryon asymmetry) and, as the oscillations have a slightly different 
initial conditions in different regions, we have an additional source of density perturbations, which adds to the inhomogeneous RH neutrino decay.

Whether the Universe becomes dominated by the scalar oscillations or not is, however, very model dependent. First, its amplitude of oscillation could be much 
smaller than $M_P$ either because of initial conditions or because the typical scale of variation for $\phi$ is much smaller than $M_P$. This happens for instance for an axion-like scalar, where the typical range of variation is the decay parameter $f$, in general much smaller than $M_P$. 
Second, its 
decay into some hidden sector or into the visible one can be fast enough to avoid the domination. In the second case one must be careful that the 
additional interactions are not so strong to induce a huge displacement of $\phi$ before RH neutrino decay. Given our complete freedom (ignorance) about
the coupling of $\phi$, it is easy to check that this can be done.

Another interesting possibility about the fate of $\phi$ would be that the scalar is still light in the present Universe and possibly observable \cite{wip}.

\section{\label{sec:conclusions}Conclusions} 
In this paper we studied a concrete realization of the inhomogeneous decay scenario \cite{Dvali:2003ar}
which is based on standard thermal leptogenesis. We considered the simplest
possibility: the SM extended with 3 hierarchical RH neutrinos. Both curvature
perturbations and the baryon asymmetry are produced through the out-of-equilibrium decay
of RH neutrinos whose parameters (mass and Yukawa couplings) are controlled by a light
scalar $\phi$.

The most interesting feature of the model is that a baryon isocurvature
mode is produced which is correlated with the adiabatic one. Though the amplitude of this
mode is model-dependent, we generically expect that detection should be very close. 
Focusing on the lightest RH neutrino, the dynamics is completely described by a single parameter
$\widetilde{m}_1$. It sets the departure from equilibrium at decay, which controls the efficiency of both
leptogenesis and the production of perturbations. The present limits
on non-Gaussianity constrain $\widetilde{m}_1$ to be in the weak wash-out (strongly
out-of-equilibrium) regime. 

We also studied the evolution of the scalar field responsible for perturbations. The
couplings of the scalar with the neutrino sector induce a motion of the scalar field
itself, which in turn induces variations in the neutrino parameters. This puts additional
constraints on the model. In the case where $\phi$ enters only in the mass of the RH
neutrinos, $\phi$ is pulled over many Planck masses during RH neutrino domination. This possibility is
therefore disfavored. On the other hand, if $\phi$ enters only in the Yukawa couplings $\lambda$,
the pull is suppressed by $\lambda^2$ so that it can be made sufficiently small.

Many generalizations of the model are possible, the simplest being an embedding in supersymmetry.
In this case the right-handed sneutrino is an additional source of baryon asymmetry. Moreover the possibility of using a sneutrino direction 
as inflaton or curvaton has been thoroughly studied \cite{Murayama:1992ua,Moroi:2002vx}. This opens a broad range of model building possibilities.

\section*{Acknowledgments}
We thank Justin Khoury and Uros Seljak for useful discussions.

\end{document}